\documentclass[11pt]{article}

\usepackage{amsmath,amssymb,amsthm}
\usepackage[numbers]{natbib}
\usepackage{graphicx}
\usepackage{booktabs}
\usepackage{geometry}
\usepackage{bm}
\usepackage{enumitem}
\usepackage{caption}
\usepackage{subcaption}
\usepackage{float}
\usepackage{mathtools}
\usepackage{xcolor}
\usepackage{tikz}
\usepackage{array}
\usepackage{xcolor}
\usepackage{tabularx}

\definecolor{deepblue}{RGB}{0,70,140}
\definecolor{tealgreen}{RGB}{0,120,110}
\definecolor{darkred}{RGB}{150,40,40}
\usepackage[
    colorlinks=true,
    linkcolor=darkred,
    citecolor=tealgreen,
    urlcolor=deepblue
]{hyperref}

\usetikzlibrary{arrows.meta,positioning,fit,shapes.geometric,calc}

\usepackage{titling}
\usepackage{array}
\preauthor{\begin{center}
    \large \lineskip .75em%
        \begin{tabular}[t]{>{\centering\arraybackslash}p{.45\textwidth}}}
        \postauthor{\end{tabular}\par\end{center}}
\makeatletter
\renewcommand\and{
  \end{tabular}%
  \hfill
  \begin{tabular}[t]{>{\centering\arraybackslash}p{.45\textwidth}}}
\makeatother

\renewcommand{\toprule}{\hline\hline}
\renewcommand{\midrule}{\hline}
\renewcommand{\bottomrule}{\hline}

\title{%
Learning effective Sargassum transport dynamics from limited drifter observations
}

\author{%
Francisco J. Beron-Vera\thanks{Corresponding author.}\\
Department of Atmospheric Sciences\\
Rosenstiel School of Marine, Atmospheric, and Earth Science\\
University of Miami\\
Miami, Florida, USA\\
\href{mailto:fberon@miami.edu}{\texttt{fberon@miami.edu}}
\and
Maria J. Olascoaga\\
Department of Ocean Sciences\\
Rosenstiel School of Marine, Atmospheric, and Earth Science\\
University of Miami\\
Miami, Florida, USA\\
\href{mailto:jolascoaga@miami.edu}{\texttt{jolascoaga@miami.edu}}
\and
Julio Morell\\
Department of Marine Sciences\\
University of Puerto Rico\\
Mayag\"uez, Puerto Rico\\
\href{mailto:julio.morell@upr.edu}{\texttt{julio.morell@upr.edu}}
\and
Edward Cruz\\
Department of Marine Sciences\\
University of Puerto Rico\\
Mayag\"uez, Puerto Rico\\
\href{mailto:edward.cruz1@upr.edu}{\texttt{edward.cruz1@upr.edu}}
}

\date{Sarted: May 10, 2026.  This version: \today.}

\begin{document}

\maketitle

\begin{abstract}
Floating-material transport is influenced by unresolved processes that are often absent from available circulation products. We develop a data-driven transport-learning framework for learning effective transport corrections from limited Lagrangian observations using physically motivated ocean--atmosphere diagnostics and finite-memory representations motivated in part by inertial-particle memory effects. The diagnostic representation is analyzed through predictive and sparse symbolic-discovery approaches under leave-one-trajectory-out validation. Applications to Sargassum-following drifters in the Puerto Rico region and the Gulf Stream show that the diagnostics contain transport-relevant information beyond the baseline circulation products. Multilayer perceptron (MLP) ensembles provide flexible predictive trajectory corrections, while Sparse Identification of Nonlinear Dynamics (SINDy) tests whether instantaneous or delayed sparse symbolic transport structure can be extracted from the diagnostics. The results differ across flow regimes: (i) in Puerto Rico, delayed sparse symbolic corrections provide modest but systematic improvement; (ii) in the Gulf Stream application, dynamically useful sparse symbolic corrections remain primarily instantaneous even though delayed predictive information persists. These results support finite-memory transport effects in coarse-grained floating-material transport while also illustrating the difficulty of obtaining stable delayed sparse symbolic closures.
\end{abstract}

\noindent\textbf{Transport models for floating material often omit unresolved processes that influence observed trajectories. Oceanic and atmospheric diagnostics extracted from limited drifter observations are analyzed within a finite-memory transport dynamical-system representation motivated in part by memory effects in inertial-particle transport. The diagnostic representation is examined through predictive trajectory correction and sparse symbolic discovery: neural-network residual prediction improves trajectory forecasts, while sparse identification tests whether interpretable symbolic transport structure can be extracted from the diagnostics. Applications to Sargassum-following drifters in the Puerto Rico region and the Gulf Stream show that delayed predictive information can persist even when stable delayed sparse symbolic closures do not emerge.}

\tableofcontents

\section{Introduction}

Large Sargassum influxes have become an important environmental, ecological, and socioeconomic challenge throughout the tropical and subtropical Atlantic, Caribbean Sea, and Gulf of Mexico \cite{Wang-etal-19, Jouanno-etal-25, Beron-etal-26-PNAS}. Anticipating transport pathways and coastal landfall locations remains difficult because available circulation products often fail to represent the effective motion of floating Sargassum aggregations.

Unlike ideal passive tracers, floating Sargassum mats may be influenced by processes beyond resolved ocean circulation. Recent work has emphasized finite-size effects, including inertial transport mechanisms and interactions among Sargassum mats \cite{Beron-Miron-20, Bonner-etal-24}. Additional unresolved contributions may arise from wave forcing, Stokes drift, unresolved submesoscale transport, coastal interactions, and other mechanisms affecting slip relative to the surrounding flow. As a result, trajectory predictions obtained solely from ocean circulation products can accumulate substantial transport errors even when the large-scale circulation is reasonably captured. These effects also raise the possibility that effective Sargassum transport may not be purely instantaneous, but may instead contain finite-memory structure.

This study is motivated by observational datasets consisting of drifters deployed directly within Sargassum mats. These drifters are intended to follow the motion of the Sargassum itself rather than ideal passive particles and therefore provide observational proxies for effective Sargassum transport. Two applications motivate the analysis: embedded-drifter observations initialized within Sargassum mats near Puerto Rico and longer trajectories in the Gulf Stream region. While the drifters were deployed within Sargassum aggregations, continuous association with the mats over the full trajectories remains to be verified. The datasets span different dynamical environments and observational characteristics but share a common goal: inferring effective transport behavior from sparse Lagrangian observations.

The limited number of available trajectories makes the problem challenging, but also raises a broader dynamical-systems question: what effective transport dynamics can be inferred from a limited collection of Sargassum-following observations?

More specifically, we ask two related questions:
\begin{enumerate}[label=(\roman*),nosep]
    \item do local oceanic and atmospheric diagnostics contain predictive transport information beyond that present in baseline circulation products? and
    \item does delayed finite-memory information provide additional predictive or sparse symbolic transport structure?
\end{enumerate}

The goal is to determine what transport structure can be recovered from sparse Lagrangian observations and how this depends on the underlying transport representation. Unlike recent efforts focused on the collective dynamics and internal evolution of Sargassum aggregations \cite{Beron-Bonner-26}, which sought mechanistic descriptions of Sargassum interactions and center-of-mass evolution using machine-learning and sparse-discovery approaches, the present work focuses on effective transport laws inferred directly from embedded Sargassum-following trajectories and on unresolved transport behavior at the observational level.

The approach is motivated in part by recent efforts combining imperfect transport models with machine-learning corrections for inertial-particle dynamics \cite{Wan-Sapsis-18}, machine-learning reconstruction of ocean surface dynamics from drifter trajectories \cite{Aksamit-etal-20}, and previous adaptations of inertial transport ideas to Sargassum dynamics \cite{Bonner-etal-24}. Inertial-particle transport frameworks based on the Maxey--Riley equation contain history-dependent effects through the Basset memory term, motivating the use of delayed finite-memory transport representations. Wan and Sapsis proposed delayed feature representations motivated by Maxey--Riley transport theory, and related ideas were subsequently adapted to Sargassum applications. Here delayed representations are examined in a different observational setting, where limited embedded-drifter trajectories are combined with physically motivated ocean--atmosphere diagnostics rather than a prescribed transport law.

We seek a reproducible pipeline for diagnosing transport mechanisms not explicitly represented in baseline circulation products. The framework combines four ingredients:
\begin{enumerate}[label=(\roman*),nosep]
    \item a baseline ocean trajectory model;
    \item physics-guided ocean--atmosphere feature construction;
    \item leave-one-trajectory-out validation; and
    \item predictive and sparse symbolic analyses using both instantaneous and delayed transport representations.
\end{enumerate}

The central objective is to investigate what transport structure is contained in instantaneous and delayed ocean--atmosphere representations constructed from sparse Lagrangian observations. In the present implementation, predictive residual corrections are represented using multilayer perceptrons (MLP and delayed MLP, dMLP; \cite{Rumelhart-etal-86}), while sparse symbolic discovery uses Sparse Identification of Nonlinear Dynamics (SINDy and delayed SINDy, dSINDy; \cite{Brunton-etal-16}) together with sequential thresholded least squares (STLSQ). The delayed SINDy formulation is of particular interest because it probes whether finite-memory transport structure admits a stable sparse symbolic representation. MLP models are used because they provide comparatively simple architectures with relatively few trainable parameters for the limited-data setting considered here, while delayed representations incorporate finite-memory information directly through the feature construction itself. Both predictive and sparse symbolic components remain replaceable without modifying the remainder of the workflow. Alternative predictive architectures include recurrent neural networks, gated recurrent units, long short-term memory networks, transformer-based models, or related approaches, while sparse identification could employ ridge-regularized STLSQ, lasso, elastic net, or Bayesian sparse regression.

\section{Learning framework}

\subsection{Pipeline overview}

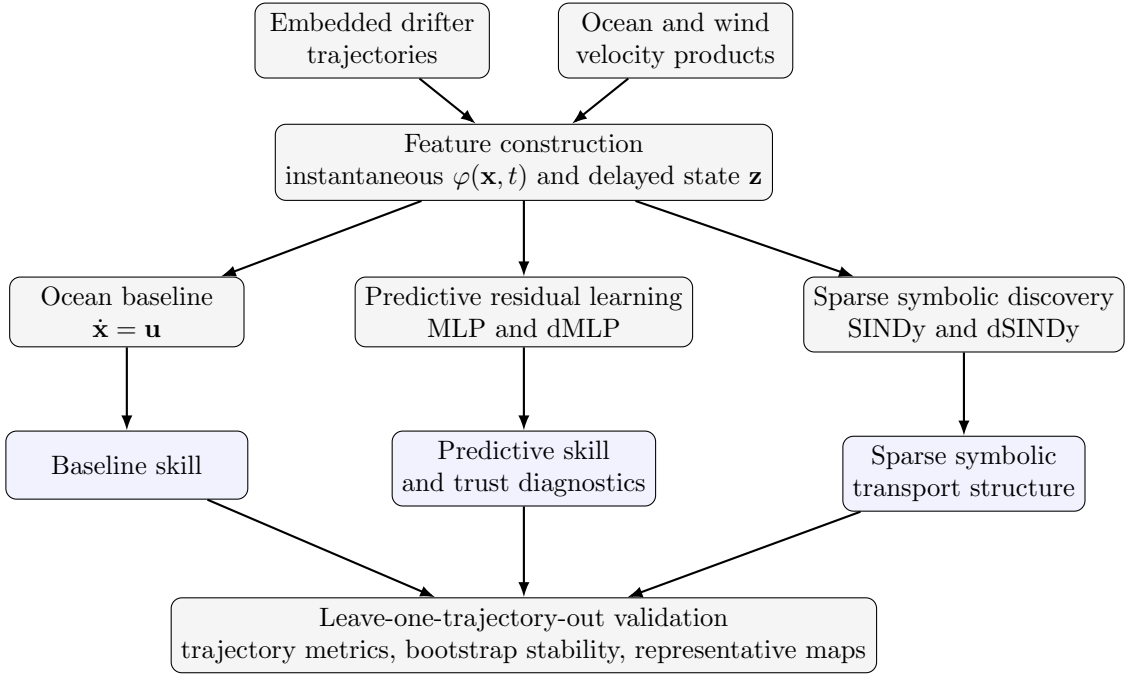
\begin{figure}[t!]
    \centering
    \begin{tikzpicture}[
        node distance=1.1cm and 0.9cm,
        box/.style={
            draw,
            rounded corners,
            align=center,
            minimum width=3.1cm,
            minimum height=0.9cm,
            font=\small,
            fill=gray!8},
        wide/.style={
            draw,
            rounded corners,
            align=center,
            minimum width=4.7cm,
            minimum height=0.9cm,
            font=\small,
            fill=gray!8},
        result/.style={
            draw,
            rounded corners,
            align=center,
            minimum width=3.2cm,
            minimum height=0.9cm,
            font=\small,
            fill=blue!5},
        arrow/.style={-{Latex[length=2mm]}, thick}
    ]

        \node[box] (data)
        {Embedded drifter\\
        trajectories};

        \node[box, right=of data] (fields)
        {Ocean and wind\\
        velocity products};

        \node[wide, below=of $(data)!0.5!(fields)$] (features)
        {Feature construction\\
        instantaneous \(\varphi(\mathbf x,t)\) and delayed state \(\mathbf z\)};

        \node[box, below left=1.0cm and 0.4cm of features] (baseline)
        {Ocean baseline\\
        \(\dot{\mathbf x}=\mathbf u\)};

        \node[box, below=1.0cm of features] (mlp)
        {Predictive residual learning\\
        MLP and dMLP};

        \node[box, below right=1.0cm and 0.4cm of features] (sindy)
        {Sparse symbolic discovery\\
        SINDy and dSINDy};

        \node[result, below=of baseline] (obase)
        {Baseline skill};

        \node[result, below=of mlp] (omlp)
        {Predictive skill\\
        and trust diagnostics};

        \node[result, below=of sindy] (osindy)
        {Sparse symbolic\\
        transport structure};

        \node[wide, below=1.2cm of omlp] (validation)
        {Leave-one-trajectory-out validation\\
        trajectory metrics, bootstrap stability, representative maps};

        \draw[arrow] (data) -- (features);
        \draw[arrow] (fields) -- (features);
        \draw[arrow] (features) -- (baseline);
        \draw[arrow] (features) -- (mlp);
        \draw[arrow] (features) -- (sindy);
        \draw[arrow] (baseline) -- (obase);
        \draw[arrow] (mlp) -- (omlp);
        \draw[arrow] (sindy) -- (osindy);
        \draw[arrow] (obase) -- (validation);
        \draw[arrow] (omlp) -- (validation);
        \draw[arrow] (osindy) -- (validation);

    \end{tikzpicture}
    \caption{
    Schematic of the learning and sparse-discovery pipeline. Instantaneous diagnostics \(\varphi(\mathbf x,t)\) and delayed finite-memory states \(\mathbf z\) are constructed from common drifter observations and ocean--atmosphere velocity products. Predictive residual learning compares instantaneous MLP and delayed MLP (dMLP) trajectory corrections, while sparse symbolic discovery compares instantaneous SINDy and delayed SINDy (dSINDy) transport closures. All models are evaluated under the same leave-one-trajectory-out protocol.
    }
    \label{fig:pipeline}
\end{figure}

The workflow is summarized in Fig.~\ref{fig:pipeline}. Embedded drifter trajectories and ocean--atmosphere velocity products are used to construct two related transport representations: an instantaneous feature vector \(\varphi(\mathbf x,t)\) and a delayed finite-memory state \(\mathbf z\). Instantaneous models use local diagnostic information only, whereas delayed models additionally incorporate finite-memory effects through lagged feature representations.

\subsection{Effective transport formulation}

We seek an effective dynamical system describing Sargassum motion using observational drifter trajectories together with baseline ocean circulation products. The baseline circulation is assumed to capture an important fraction of the large-scale dynamics, while the learned component represents unresolved contributions. This viewpoint is motivated in part by recent work combining imperfect transport models with machine-learning corrections in inertial particle dynamics \cite{Wan-Sapsis-18}.

The construction is also motivated by oceanographic adaptations of the Maxey--Riley equation \cite{Maxey-Riley-83} for finite-size, massive (inertial) transport, including theoretical, observational, laboratory, and geophysical studies of inertial effects in floating matter and Sargassum aggregations \cite{Beron-etal-15, Beron-etal-19-PoF, Olascoaga-etal-20, Miron-etal-20-GRL, Miron-etal-20-PoF, Beron-21-ND, Beron-Miron-20, Olascoaga-etal-23, Bonner-etal-24, Beron-24-POFb}. In these studies, the full inertial dynamics involve coupled slow and fast variables, while reduced slow-manifold equations depend on quantities beyond the resolved velocity field, including atmospheric forcing, acceleration, vorticity, and memory effects associated with the Basset history term. These considerations motivate the use of delayed finite-memory states in the present setting. Sargassum motion may additionally involve unresolved wave effects, Stokes drift, unresolved submesoscale transport, coastal interactions, and systematic forcing or model biases. The goal is not to impose a specific inertial transport law, but to use these ideas as guidance for constructing effective transport states.

Two related closures are considered. The first uses an instantaneous state,
\begin{equation}
    \dot{\mathbf x}
    =
    \mathbf u(\mathbf x,t)
    +
    \alpha(\varphi)
    \mathbf G_\beta(\varphi),
    \label{eq:model_markov}
\end{equation}
where \(\varphi=\varphi(\mathbf x,t)\) denotes an instantaneous feature vector constructed from local ocean--atmosphere diagnostics. The second introduces delayed finite-memory states,
\begin{equation}
    \dot{\mathbf x}
    =
    \mathbf u(\mathbf x,t)
    +
    \alpha(\mathbf z)
    \mathbf G_\beta(\mathbf z),
    \label{eq:model_delayed}
\end{equation}
where \(\mathbf z\) augments the instantaneous diagnostics with delayed variables. In both formulations, \(\mathbf u\) denotes the baseline ocean velocity field, \(\mathbf G_\beta\) denotes a learned residual correction parameterized by trainable coefficients \(\beta\), and \(\alpha\) denotes a trust factor suppressing corrections in poorly constrained regions of state space.

The learned residual is treated as an effective unresolved contribution,
\begin{equation}
    \mathbf u_{\rm eff}
    =
    \mathbf u
    +
    \mathbf u_{\rm res},
\end{equation}
with
\begin{equation}
    \begin{aligned}
        \mathbf u_{\rm res}
        =\;&
        \text{finite-size/inertial effects}
        \\
        &+
        \text{wave and Stokes-drift effects}
        +
        \text{unresolved submesoscale transport}
        \\
        &+
        \text{model bias}
        +
        \cdots.
    \end{aligned}
\end{equation}

The learned correction therefore augments rather than replaces the baseline circulation model. Instantaneous closures depend only on local diagnostics, whereas delayed closures additionally incorporate lagged finite-memory information.

\subsection{Physics-informed diagnostics}

Guided by the inertial and windage considerations above, instantaneous diagnostics are constructed from local oceanic and atmospheric quantities,
\begin{equation}
    \varphi(\mathbf x,t)
    =
    \big\{
    \mathbf u,
    \mathbf u_a,
    D_t\mathbf u,
    D_t\mathbf u_a,
    \omega,\omega_a,
    \nabla\!\cdot\mathbf u,
    \nabla\!\cdot\mathbf u_a
    \big\},
    \label{eq:phi}
\end{equation}
where \(\mathbf u=(u,v)\) denotes ocean velocity and \(\mathbf u_a=(u_a,v_a)\) denotes atmospheric velocity. The material derivative is written compactly as
\begin{equation}
    D_t
    :=
    \frac{D}{Dt}
    =
    \partial_t
    +
    (\mathbf u\cdot\nabla).
\end{equation}
Vertical vorticity is defined by
\begin{equation}
    \omega
    =
    \partial_x v
    -
    \partial_y u.
\end{equation}
Analogous definitions are used for atmospheric quantities.

Two transport states are considered. The instantaneous state uses only the local diagnostics,
\begin{equation}
    \mathbf z_n
    =
    \varphi_n,
    \qquad
    \varphi_n := \varphi(\mathbf x(t_n),t_n)
    \label{eq:instantaneouslibrary}
\end{equation}
corresponding to an effective Markovian closure. The delayed state augments these diagnostics with finite-memory information,
\begin{equation}
    \mathbf z_n
    =
    \left\{
    \varphi_n,
    \varphi_{n-1},
    \ldots,
    \varphi_{n-p}
    \right\},
    \label{eq:delayedlibrary}
\end{equation}
where \(p\) denotes the number of retained delays. Thus \(p=0\) recovers the instantaneous state, while \(p>0\) introduces delayed short-memory structure. In the present delayed implementations, \(p=2\).

The delayed variables represent the recent Lagrangian environment sampled by the Sargassum mat through lagged ocean--atmosphere diagnostics. The construction is motivated partly by reduced inertial and windage ideas, although no specific transport law is imposed.

\subsection{MLP residual model}

\subsubsection{Model definition}

In the present implementation, the correction map \(\mathbf G_\beta\) is represented by a feed-forward neural network (FNN), specifically a multilayer perceptron (MLP). Both instantaneous and delayed residual models are considered. Instantaneous MLP uses the local feature vector \(\varphi(\mathbf x,t)\), whereas delayed MLP (dMLP) uses the finite-memory state \(\mathbf z\), allowing the correction to depend on the recent transport history sampled by the trajectory.

Memory enters explicitly through the delayed state construction rather than through recurrent hidden states. This permits the use of relatively simple feed-forward networks instead of more complicated recurrent architectures such as long short-term memory (LSTM) networks or gated recurrent units (GRUs). Given the limited number of available trajectories and sparse sampling of state space, MLP models provide flexible nonlinear corrections with comparatively few trainable parameters. Other predictive architectures could nevertheless be incorporated without modifying the remainder of the pipeline.

For a network with $L$ hidden layers,
\begin{equation}
    \mathbf G_\beta(\mathbf z)
    =
    \mathbf W_{L+1}
    \sigma\!\Big(
    \mathbf W_L
    \sigma\!\big(
    \cdots
    \sigma(
    \mathbf W_1\mathbf z+\mathbf b_1)
    \cdots
    \big)
    +
    \mathbf b_L
    \Big)
    +
    \mathbf b_{L+1},
\end{equation}
where $\sigma$ denotes a nonlinear activation function and
\begin{equation}
    \beta
    =
    \{
    \mathbf W_1,\mathbf b_1,
    \ldots,
    \mathbf W_{L+1},\mathbf b_{L+1}
    \}
\end{equation}
collects the trainable weights and biases.

The parameters are determined from residual velocities through
\begin{equation}
    \beta^\ast
    =
    \underset{\beta}{\operatorname{arg\,min}}
    \;
    \frac1N
    \sum_{n=1}^{N}
    \left\|
    \dot{\mathbf x}_n
    -
    \mathbf u_n
    -
    \mathbf G_\beta(\mathbf z(t_n))
    \right\|^2.
\end{equation}

The optimization problem is nonconvex and is solved through backpropagation and stochastic gradient-based optimization. Different random initializations and optimization paths can therefore lead to distinct trained models, motivating the ensemble strategy introduced below. The learned network represents the residual velocity correction required to reproduce the observed Sargassum-following transport.

\subsubsection{Leave-one-trajectory-out ensemble}

Because only a small number of trajectories are available, validation uses a leave-one-trajectory-out (LOTO) strategy. For each experiment, one trajectory is withheld entirely from training and reserved for testing. The remaining trajectories are then used to train ensembles of instantaneous MLP and delayed MLP (dMLP) residual models. Ensemble methods are widely used to improve robustness and quantify predictive uncertainty \cite{Lakshminarayanan-etal-17}. Here each ensemble member uses identical training data and architecture but begins from a different random initialization, thereby sampling distinct optimization pathways and local minima.

During testing, all ensemble members are integrated independently from the same initial condition, producing a collection of corrected trajectories
\[
\mathbf x_{\rm MLP}^{(1)},
\ldots,
\mathbf x_{\rm MLP}^{(K)}.
\]
The deployed prediction is the ensemble-mean trajectory,
\begin{equation}
    \overline{\mathbf x}_{\rm MLP}(t)
    =
    \frac{1}{K}
    \sum_{k=1}^{K}
    \mathbf x_{\rm MLP}^{(k)}(t),
\end{equation}
where \(K\) denotes the ensemble size. The same ensemble procedure is applied to both instantaneous and delayed residual models.

\subsubsection{Trust modulation}

To reduce corrections in regions of limited support, we introduce a trust mechanism motivated by ensemble-based uncertainty ideas \cite{Lakshminarayanan-etal-17}. The approach uses disagreement among ensemble members as a proxy for prediction uncertainty.

Ensemble spread is quantified through the sample standard deviation of ensemble trajectory predictions,
\begin{equation}
    \sigma(\mathbf z)
    :=
    \left(
    \frac{1}{K-1}
    \sum_{k=1}^{K}
    \left\|
    \mathbf x_{\rm MLP}^{(k)}
    -
    \overline{\mathbf x}_{\rm MLP}
    \right\|^2
    \right)^{1/2}.
\end{equation}
The trust factor is then defined as
\begin{equation}
    \alpha(\mathbf z)
    :=
    \exp
    \left(
    -
    \frac{\sigma(\mathbf z)^2}
    {\sigma_0^2}
    \right).
\end{equation}
Small values of \(\sigma_0\) suppress corrections aggressively, while \(\sigma_0=\infty\) corresponds to no trust suppression. The trust factor does not guarantee predictive accuracy: ensemble members may agree on an incorrect correction. It nevertheless acts as a safeguard against unconstrained extrapolation and provides a diagnostic measure of prediction confidence.

Instantaneous and delayed models can exhibit different ensemble-spread behavior across transport regimes. In some cases, local diagnostics alone already produce stable ensemble predictions. In others, delayed states reduce ensemble spread or improve predictive stability, indicating that recent transport history contributes useful predictive structure.

\subsection{Sparse discovery}

\subsubsection{SINDy formulation}

SINDy seeks parsimonious symbolic descriptions by representing a target quantity as a sparse linear combination of candidate functions. Here sparse symbolic regression is applied to the residual transport contribution rather than to the full dynamics. The deployed model is written as
\begin{equation}
    \dot{\mathbf x}
    =
    \mathbf u(\mathbf x,t)
    +
    \alpha(\mathbf z)
    \Theta(\mathbf z)
    \boldsymbol\xi,
    \label{eq:sindymodel}
\end{equation}
where $\Theta(\mathbf z)$ denotes a library of candidate functions constructed from the feature state $\mathbf z$, and $\boldsymbol\xi$ contains sparse coefficients determining the active correction terms. In analogy with the neural-network parameter vector $\beta$, the sparse coefficient vector $\boldsymbol\xi$ plays the role of the learned symbolic closure parameters. The trust factor $\alpha(\mathbf z)$ is retained during deployment and suppresses corrections in regions of low confidence.

For instantaneous SINDy, \(\mathbf z=\varphi(\mathbf x,t)\) and the candidate library depends only on local diagnostics. Delayed SINDy (dSINDy) instead augments the library with delayed states from Eq.~\eqref{eq:delayedlibrary}, allowing sparse symbolic corrections to depend explicitly on finite-memory structure. In the present implementation, candidate terms consist of oceanic and atmospheric diagnostics selected from the physically motivated feature set introduced above.

The sparse symbolic-regression problem is written as
\begin{equation}
    \boldsymbol\xi^\ast
    =
    \underset{\boldsymbol\xi}{\operatorname{arg\,min}}
    \left\{
    \left\|
    \mathbf r
    -
    \Theta(\mathbf z)
    \boldsymbol\xi
    \right\|_2^2
    +
    \lambda
    \|\boldsymbol\xi\|_0
    \right\},
\end{equation}
where
\begin{equation}
    \mathbf r
    =
    \dot{\mathbf x}
    -
    \mathbf u
\end{equation}
denotes residual velocity and \(\|\cdot\|_0\) counts active terms. The optimization balances residual error against model complexity. Direct minimization of the \(L_0\) penalty is generally combinatorial and NP-hard because it requires selecting active terms among many possible candidate subsets.

The current implementation uses STLSQ (sequential thresholded least squares), the sparse-regression strategy originally introduced with SINDy, as an efficient approximation to the formal sparse optimization problem. STLSQ alternates between least-squares fitting and thresholding steps that remove coefficients below a prescribed magnitude, progressively eliminating weak terms and promoting parsimonious symbolic models. The sparse symbolic component remains modular: regularized alternatives including ridge-STLSQ, lasso, elastic net, and Bayesian sparse regression can be substituted without changing the remainder of the pipeline.

Instantaneous SINDy depends only on local diagnostics, whereas delayed SINDy incorporates lagged finite-memory states directly into the symbolic closure.

\subsubsection{Bootstrap stability}

Sparse symbolic-transport performance and sparse symbolic-structure stability are assessed separately. LOTO (leave-one-trajectory-out) experiments determine whether the learned correction improves out-of-sample trajectories, whereas bootstrap resampling with replacement evaluates the robustness of the discovered sparse symbolic structure under perturbations of the training data.

For a candidate library term \(j\) in \(\Theta(\mathbf z)\), the bootstrap active frequency is defined as
\begin{equation}
    f_j
    =
    \frac{
    \#\{\text{bootstrap fits in which term }j\text{ is active}\}
    }{
    \#\{\text{bootstrap fits}\}
    }.
\end{equation}
Large values of \(f_j\) indicate that sparse symbolic regression repeatedly selects the same term across resampled datasets. Median active frequency measures overall symbolic-structure stability, while the largest values of \(f_j\) identify the most persistent terms across bootstrap realizations.

Bootstrap stability and trajectory prediction need not coincide. Some sparse symbolic structures persist across resampled datasets while producing little improvement in out-of-sample trajectory evolution. In other cases, stable sparse structure and trajectory improvement occur simultaneously, supporting recurrent transport mechanisms associated with the chosen state representation.

\subsection{Validation metrics}

Model performance is evaluated on withheld trajectories by comparing predicted and observed positions at identical observation times $t_n$. Errors are therefore evaluated directly in time-synchronized trajectory space. Let $\mathbf x_{\rm obs}(t_n)$ denote the observed trajectory and let $\mathbf x_m(t_n)$ denote the prediction generated by model $m$, where $m$ may correspond to the ocean baseline, MLP ensemble, or SINDy closure. The root-mean-square error is defined as
\begin{equation}
    E_{\rm RMS}(m)
    =
    \left(
    \frac1N
    \sum_{n=1}^{N}
    \left\|
    \mathbf x_m(t_n)
    -
    \mathbf x_{\rm obs}(t_n)
    \right\|^2
    \right)^{1/2}.
\end{equation}
Skill is reported relative to the ocean baseline through
\begin{equation}
    R_m
    =
    \frac{
    E_{\rm RMS}(m)
    }{
    E_{\rm RMS}(\mathrm{ocean})
    }.
\end{equation}
Thus $R_m<1$ indicates improvement over the ocean-model trajectory, whereas $R_m>1$ indicates degradation. Mean and final-time trajectory errors are defined analogously.

\subsection{Trajectory integration with delayed representations}

For trajectory prediction, the learned correction is coupled to the baseline ocean velocity field and integrated forward from the observed initial condition. Delayed models evolve on augmented finite-memory states constructed from recent transport history. At the continuous level, the resulting dynamics may be viewed schematically as a finite-memory system of the form
\begin{equation}
    \dot{\mathbf x}(t)
    =
    \mathbf f\!\big(
    \mathbf x(t),
    \mathbf x(t-\tau_1),
    \ldots,
    \mathbf x(t-\tau_p),
    t
    \big),
    \qquad
    0 < \tau_1 < \cdots < \tau_p,
\end{equation}
supplemented by a history condition
\begin{equation}
    \mathbf x(t)
    =
    \mathbf y(t),
    \qquad
    t\in[t_0-\tau_p,t_0].
\end{equation}

In the present formulation, delayed dependence enters through diagnostic coordinates rather than directly through delayed positions. The delayed state consists of a finite collection of previous ocean--atmosphere diagnostics, which are evolved recursively together with the model trajectory during integration rather than prescribed from withheld observations.

For delayed MLP models, the memory is encoded implicitly through the delayed state itself:
\[
\text{current delayed state}
\rightarrow
\text{dMLP}
\rightarrow
\text{correction}
\rightarrow
\text{next state}.
\]
The neural network therefore acts as an instantaneous nonlinear map on the augmented delayed state space.

For delayed SINDy models, the delayed variables enter explicitly into the symbolic transport closure through sparse combinations of delayed diagnostic coordinates. Delayed SINDy therefore defines an explicit finite-memory transport law in terms of delayed diagnostic states.

Instantaneous MLP and SINDy correspond to the Markovian limit \(p=0\), where the delayed state reduces to the local diagnostic vector.

\section{Applications}

\subsection{Puerto Rico drifters}

The Puerto Rico application considers drifters deployed on Sargassum mats off southern Puerto Rico as part of observational efforts by the Caribbean Coastal Ocean Observing System (CARICOOS). Ocean baseline trajectories are obtained from a CARICOOS regional implementation of the Finite-Volume Community Ocean Model (FVCOM; \cite{Chen-etal-03}) for the Puerto Rico domain, while atmospheric velocity diagnostics are derived from the European Centre for Medium-Range Weather Forecasts (ECMWF) ERA5 reanalysis \cite{Hersbach-etal-20}. These ocean and atmosphere products provide the reference fields from which the instantaneous and delayed diagnostic states and learned transport corrections are constructed.

The Puerto Rico trajectories require basic quality control before delayed analysis. Observation intervals are irregular, ranging from approximately 10 minutes to 1 hour with occasional larger gaps, and one trajectory exhibits unphysical land-crossing behavior. To obtain physically interpretable delayed states, the diagnostics are constructed using fixed-$\Delta t$ interpolation with $\Delta t=1$ hr together with segmentation across large temporal gaps, and the corrupted trajectory segment is excluded from the final analysis.

Table~\ref{tab:modelsummary} summarizes the LOTO results for the Puerto Rico application. The updated comparison clarifies the role of delayed predictive structure in Puerto Rico. The instantaneous MLP already provides substantial improvement over the baseline circulation model, with median trajectory-wise RMS ratio \(R_{\mathrm{MLP}}\approx0.76\), median final-position error ratio of approximately \(0.73\), and improvement for 8 of the 12 retained trajectories. The delayed MLP also improves 8 of the 12 trajectories, but with weaker aggregate RMS improvement, \(R_{\mathrm{dMLP}}\approx0.90\), and nearly neutral aggregate final-position behavior, with final-position error ratio approximately equal to one. The Puerto Rico results therefore show that delayed memory is not required for substantial predictive improvement. The instantaneous MLP already provides a strong local predictive closure, while delayed MLP examines the effect of short-memory transport states on trajectory prediction and stability.

\begin{table}[t!]
   \centering
   \small
   \begin{tabularx}{\textwidth}{lcccccX}
       \toprule
       Model & Rep. & Median RMS & Improved & Final & Role \\
             &      & ratio       & traj.    & ratio &      \\
       \midrule
       Ocean  & baseline & 1.000 & --   & 1.000 &
       reference baseline \\

       MLP    & instant. & 0.755 & 8/12 & 0.729 &
       predictive Markovian closure \\

       dMLP   & delayed & 0.898 & 8/12 & 1.003 &
       predictive finite-memory closure \\

       SINDy  & instant. & 0.988 & --   & 1.000 &
       sparse symbolic local correction \\

       dSINDy & delayed & 0.961 & --   & 0.950 &
       sparse symbolic delayed closure \\
       \bottomrule
   \end{tabularx}
   \caption{
   Puerto Rico leave-one-trajectory-out comparison after excluding trajectory 4. Values below one indicate improvement relative to the corresponding ocean baseline. MLP and dMLP denote instantaneous and delayed predictive residual closures, respectively, while SINDy and dSINDy denote instantaneous and delayed sparse symbolic closures. The delayed representation uses a fixed-$\Delta t$ formulation with two delayed lags corresponding to 1--2 hr memory.
   }
   \label{tab:modelsummary}
\end{table}

Local instantaneous diagnostics already contain enough structure for neural residual models to improve many withheld trajectories. Delayed states are introduced to examine whether short-time transport memory yields additional predictive improvement or persistent delayed symbolic structure.

\begin{figure}[t!]
    \centering
    \includegraphics[width=\textwidth]{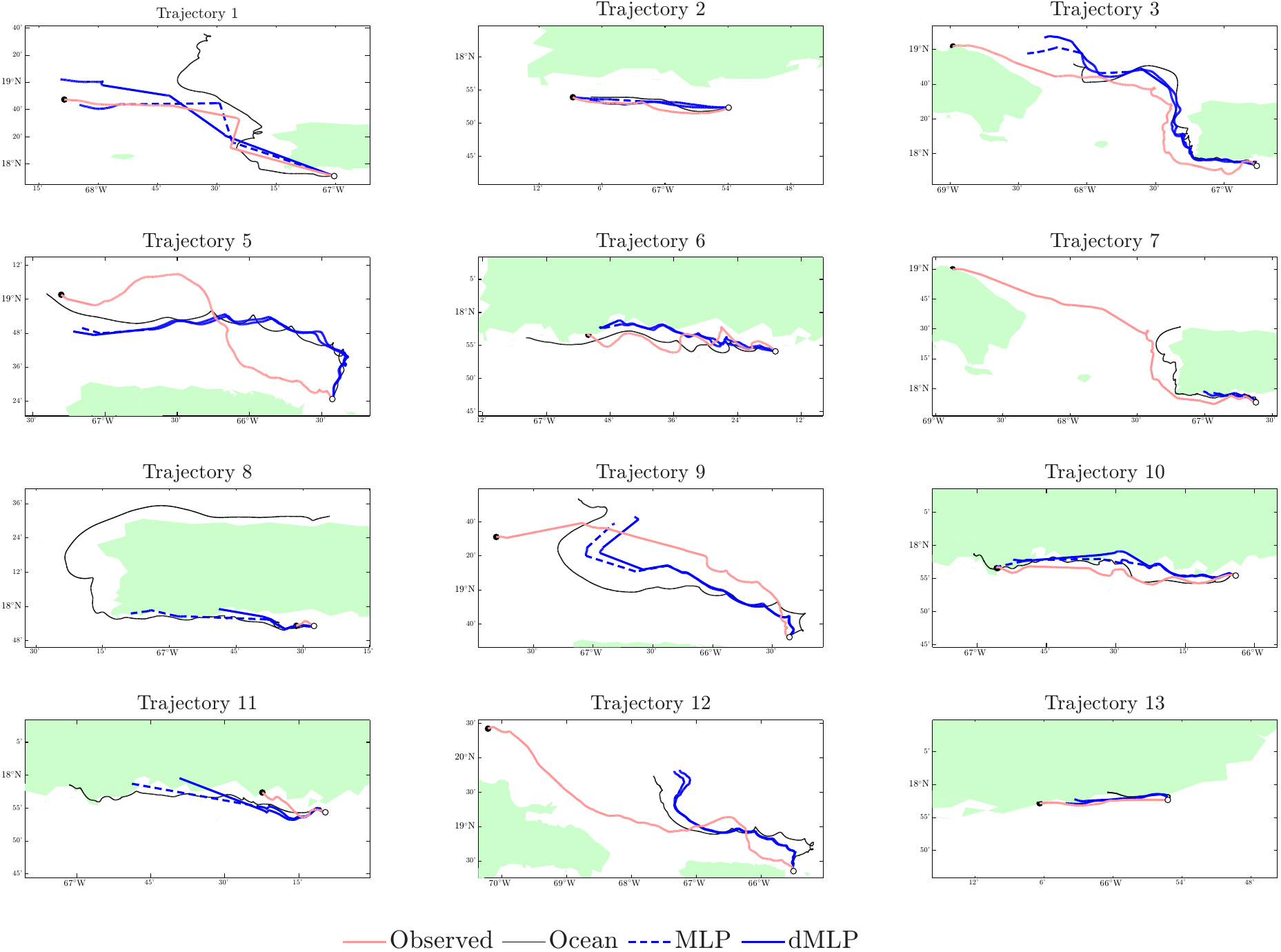}
    \caption{
    Individual leave-one-trajectory-out MLP trajectory comparisons for the Puerto Rico application. Observed ``Sargassum-following'' drifter trajectories are compared with the baseline ocean trajectory, the instantaneous MLP correction, and the delayed MLP correction. The instantaneous MLP is shown as a dashed blue curve and the delayed MLP as a solid blue curve. Filled circles denote trajectory initiation and open circles denote observed terminal locations.
    }
    \label{fig:loto_individual}
\end{figure}

Figure~\ref{fig:loto_individual} shows individual LOTO trajectory comparisons for both predictive models. Observations correspond to ``Sargassum-following'' drifter trajectories, ocean trajectories are obtained from the baseline circulation product, dashed blue curves denote the instantaneous MLP, and solid blue curves denote the delayed MLP. Aggregate statistics partially obscure substantial trajectory-to-trajectory variability. Several withheld trajectories are substantially improved by the instantaneous MLP alone, whereas others remain difficult for both MLP and dMLP. The delayed MLP acts primarily as a finite-memory extension of an already effective predictive closure rather than as a uniformly superior model.

\begin{figure}[t!]
    \centering
    \includegraphics[width=.75\textwidth]{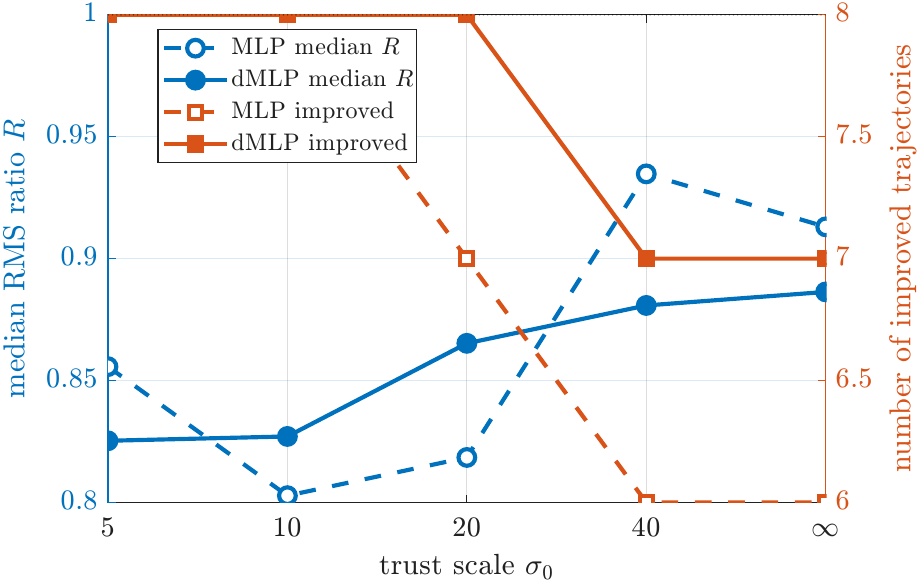}
    \caption{
    Trust-scale sensitivity for the instantaneous MLP and delayed MLP ensembles in the Puerto Rico application. Curves show the median trajectory RMS ratio and the number of withheld trajectories improved relative to the ocean baseline as functions of the trust scale $\sigma_0$. The instantaneous MLP is shown with open/dashed markers and the delayed MLP with filled/solid markers.
    }
    \label{fig:sigma_sweep}
\end{figure}

The predictive corrections are modulated through an ensemble-based trust factor intended to suppress extrapolative corrections in poorly constrained regions of state space. Figure~\ref{fig:sigma_sweep} summarizes sensitivity to the trust scale $\sigma_0$ for both MLP and dMLP. Small values suppress learned corrections strongly and keep trajectories close to the ocean baseline, whereas larger values permit stronger residual corrections. Trust modulation affects both instantaneous and delayed predictive models. The instantaneous MLP achieves the strongest median RMS improvement in the Puerto Rico case, while the delayed MLP remains a useful finite-memory comparison rather than the dominant predictive model.

\begin{figure}[t!]
    \centering
    \includegraphics[width=.75\textwidth]{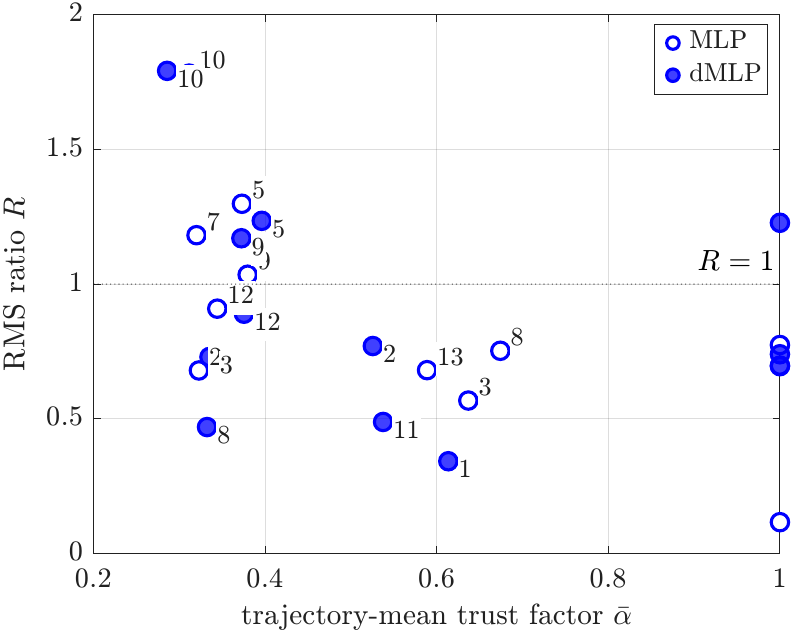}
    \caption{
    Trust--performance diagnostic for the instantaneous MLP and delayed MLP ensembles in the Puerto Rico application. Each point corresponds to one withheld trajectory. The horizontal axis is the trajectory-mean trust factor $\bar\alpha$ and the vertical axis is the trajectory RMS ratio relative to the ocean baseline. Open circles denote the instantaneous MLP and filled circles denote the delayed MLP.
    }
    \label{fig:trust_scatter}
\end{figure}

Figure~\ref{fig:trust_scatter} compares trajectory performance with the trajectory-mean trust factor. Open circles denote the instantaneous MLP and filled circles denote the delayed MLP. Ensemble spread provides a useful uncertainty proxy, but high ensemble agreement is not sufficient for accuracy. Some strongly improved trajectories occur at moderate trust levels, while several degraded cases are associated with stronger suppression. The trust factor acts primarily as an extrapolation-control diagnostic rather than a direct error estimator.

The sparse symbolic results differ substantially from the predictive neural models. Instantaneous SINDy remains nearly neutral in Puerto Rico, whereas delayed SINDy provides modest but systematic improvement together with recurrent delayed structure under bootstrap resampling. The Puerto Rico application therefore highlights dSINDy as the dominant sparse symbolic closure.

To obtain physically interpretable delayed structure, delayed states are constructed using fixed-$\Delta t$ interpolation with sampling interval $\Delta t=1$ hr together with treatment of large temporal gaps before delayed sparse symbolic regression. The delayed state consequently takes the form
\begin{equation}
    \mathbf z_n
    =
    \{\varphi_n,\varphi_{n-1},\varphi_{n-2}\},
\end{equation}
so that the retained delayed terms correspond explicitly to 1-hour and 2-hour memory. The corresponding delayed sparse symbolic transport correction can be written as
\begin{align}
    \dot x_n
    &=
    u_n
    +
    \widetilde G_x(\mathbf z_n), \\
    \dot y_n
    &=
    v_n
    +
    \widetilde G_y(\mathbf z_n).
\end{align}
A representative sparse symbolic structure inferred from the bootstrap analysis is
\begin{align}
    \widetilde G_x(\mathbf z_n)
    \approx\;&
    -u_n
    +
    0.2\,v_n
    +
    0.2\,u_{n-1}
    +
    0.4\,v_{n-1}
    \nonumber\\
    &+
    0.2\,u_{n-2}
    -
    0.3\,v_{n-2}
    +
    0.03\,(u_a)_{n-1},
    \\
    \widetilde G_y(\mathbf z_n)
    \approx\;&
    -0.8\,v_n
    -
    0.4\,u_{n-1}
    -
    0.5\,v_{n-1}
    \nonumber\\
    &+
    0.3\,u_{n-2}
    +
    0.4\,v_{n-2}
    +
    0.04\,(v_a)_n
    +
    0.04\,(D_tv)_n.
\end{align}
The displayed coefficients correspond to the most persistent delayed symbolic structures identified across bootstrap realizations; lower-frequency terms are omitted for clarity. The learned sparse symbolic corrections represent effective delayed transport closures inferred from the diagnostic states rather than direct reductions of inertial-particle transport equations.

Under this fixed-$\Delta t$ reinterpretation, delayed SINDy produces modest but systematic improvement relative to the ocean baseline while remaining more conservative than the predictive neural closures. Table~\ref{tab:modelsummary} shows that the delayed sparse symbolic closure achieves aggregate trajectory ratios below one, with the best performance obtained for a comparatively small global correction amplitude $\alpha\approx0.05$. Only a modest delayed symbolic correction is required to improve the baseline trajectories.

The bootstrap analysis further indicates that the delayed library contains persistent recurrent structure rather than a single rigid universal equation. Median active frequencies are approximately $0.46$ for the $x$ equation and $0.47$ for the $y$ equation. Table~\ref{tab:bootstrap} summarizes the most persistent delayed terms identified across bootstrap realizations. The retained sparse symbolic structures are dominated by ocean velocity memory, atmospheric forcing, material acceleration, and short-time delayed velocity coupling across both 1-hour and 2-hour lags. The recurrent delayed dependencies indicate persistent short-memory structure across bootstrap realizations, even though no single universal symbolic equation emerges across all withheld trajectories.

\begin{table}[t!]
    \centering
    \begin{tabular}{lclc}
    \toprule
    $x$-equation term & Frequency & $y$-equation term & Frequency \\
    \midrule
    $u$ & 0.996 & $v$ & 0.993 \\
    $(v_a)_{n-2}$ & 0.858 & $v_a$ & 0.985 \\
    $u_{n-2}$ & 0.820 & $u_{n-2}$ & 0.933 \\
    $v$ & 0.801 & $v_{n-1}$ & 0.799 \\
    $u_{n-1}$ & 0.785 & $D_tv$ & 0.865 \\
    $v_{n-1}$ & 0.764 & $(v_a)_{n-1}$ & 0.826 \\
    $D_tu_a$ & 0.638 & $u_{n-1}$ & 0.772 \\
    $v_{n-2}$ & 0.638 & $(v_a)_{n-2}$ & 0.703 \\
    $\omega_a$ & 0.597 & $\nabla\cdot \mathbf u$ & 0.669 \\
    $u_a$ & 0.586 & $(D_tu)_{n-1}$ & 0.650 \\
    \bottomrule
    \end{tabular}
    \caption{Most persistent delayed SINDy terms from fixed-$\Delta t$ bootstrap resampling. Subscripts denote fixed-time delays relative to sampling index $n$ with $\Delta t=1$ hr, so that $n-1$ and $n-2$ correspond to 1-hour and 2-hour lags, respectively. Frequencies denote the fraction of bootstrap fits in which a term remains active.}
    \label{tab:bootstrap}
\end{table}

\begin{figure}[t!]
    \centering
    \includegraphics[width=\textwidth]{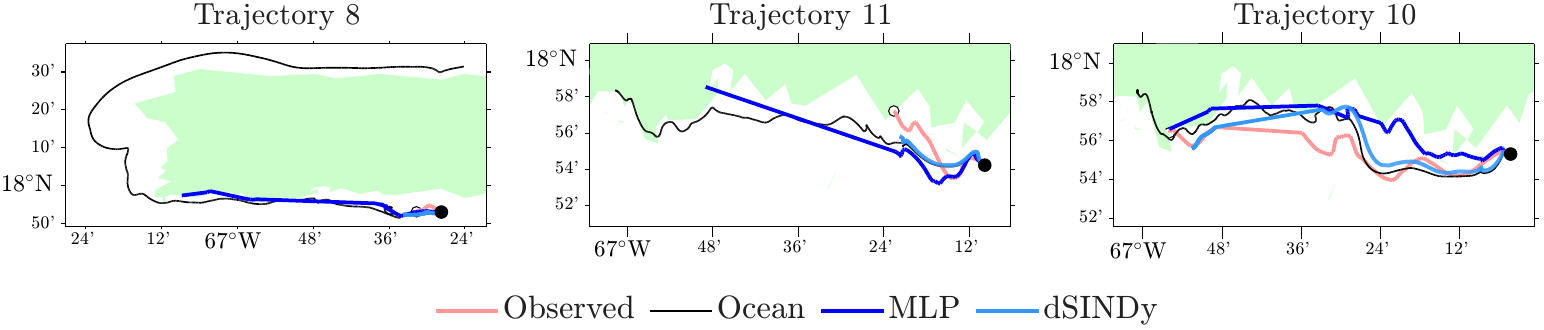}
    \caption{
    Representative leave-one-trajectory-out trajectory comparisons for the Puerto Rico application. Observed ``Sargassum-following'' drifter trajectories are compared with the ocean baseline, the instantaneous MLP correction, and the delayed sparse symbolic dSINDy correction. Filled circles and open circles indicate trajectory start and end points.
    }
    \label{fig:trajectory_examples}
\end{figure}

Figure~\ref{fig:trajectory_examples} illustrates representative withheld trajectories used to compare the predictive and sparse symbolic models. For the Puerto Rico application, the predictive model shown is the instantaneous MLP, whereas the sparse symbolic model is dSINDy. The examples highlight the different behavior of the predictive and sparse symbolic closures. MLP provides stronger trajectory-level corrections, while dSINDy remains smoother and more conservative with explicit finite-memory structure. Trajectory 8 provides the clearest delayed symbolic correction example, with dSINDy giving the cleanest correction after substantial drift in the ocean baseline. Trajectory 11 highlights the stronger predictive capability of the MLP branch, whereas trajectory 10 illustrates a more balanced regime in which the MLP becomes comparatively oscillatory while dSINDy remains smoother.

The Puerto Rico experiments show that instantaneous diagnostics already contain substantial predictive structure. A simple residual MLP improves many withheld trajectories without requiring delayed memory. Delayed states nevertheless remain dynamically meaningful through the sparse symbolic results: dSINDy identifies weak but recurrent delayed structure associated with short-time velocity memory and ocean--atmosphere transport effects.

\subsection{Gulf Stream drifters}

The Gulf Stream application provides a more stringent test because the drifter trajectories sample a dynamically energetic western-boundary-current jet together with adjacent recirculatory regimes. The drifters were deployed in Sargassum mats within the Gulf Stream off southeastern Florida. Ocean baseline trajectories are obtained from Archiving, Validation, and Interpretation of Satellite Oceanographic data (AVISO) altimetry-derived geostrophic velocity products \cite{LeTraon-etal-98}, while atmospheric velocity diagnostics are obtained from the European Centre for Medium-Range Weather Forecasts (ECMWF) ERA5 reanalysis, as in the Puerto Rico application.

The Gulf Stream trajectories require more substantial quality control before learning because of large temporal gaps, intermittently inactive ocean baselines, and difficult recirculatory regimes. The retained trajectory segments are therefore selected to ensure physically reasonable continuous evolution together with dynamically active AVISO baseline trajectories over the same time intervals. Delayed states are constructed using fixed-$\Delta t$ interpolation with $\Delta t=6$ hr, and all learned corrections are evaluated against AVISO ocean trajectories integrated over the same time vectors.

Table~\ref{tab:modelsummary_gulfstream} summarizes the LOTO results for the Gulf Stream application. The Gulf Stream experiments support an interpretation broadly consistent with the Puerto Rico case. Instantaneous diagnostics already contain substantial predictive structure, with median trajectory-wise RMS ratio \(R_{\mathrm{MLP}}\approx0.47\) and median final-position error ratio approximately equal to \(0.39\). Delayed MLP corrections can further improve some coherent jet-following regimes, but their benefits are more selective and trajectory-dependent. Delayed states are therefore not introduced because an instantaneous learned closure fails. Instead, they allow the role of short-memory transport structure to be examined within the same dynamical setting.

\begin{table}[t!]
   \centering
   \small
   \begin{tabularx}{\textwidth}{lcccccX}
       \toprule
       Model & Rep. & Median RMS & Improved & Final & Role \\
             &      & ratio       & traj.    & ratio &      \\
       \midrule
       Ocean  & baseline & 1.000 & --    & 1.000 &
       reference baseline \\

       MLP    & instant. & 0.469 & 22/26 & 0.388 &
       predictive Markovian closure \\

       dMLP   & delayed & 0.534 & 22/26 & 1.473 &
       predictive finite-memory closure \\

       SINDy  & instant. & 0.894 & --    & 0.952 &
       sparse symbolic local correction \\

       dSINDy & delayed & 1.000 & --    & 1.000 &
       sparse symbolic delayed closure \\
       \bottomrule
   \end{tabularx}

   \caption{
   Gulf Stream leave-one-trajectory-out comparison. Values below one indicate improvement relative to the corresponding AVISO ocean baseline. MLP and dMLP denote instantaneous and delayed predictive residual closures, respectively, while SINDy and dSINDy denote instantaneous and delayed sparse symbolic closures. Delayed representations use a fixed-$\Delta t$ formulation with $\Delta t=6$ hr.
   }

   \label{tab:modelsummary_gulfstream}
\end{table}

Local instantaneous diagnostics already contain enough structure for neural residual models to improve most withheld trajectories. Delayed states are introduced to examine whether finite-memory transport structure yields additional predictive improvement or persistent delayed symbolic structure in coherent Gulf Stream transport regimes.

\begin{figure}[t!]
    \centering
    \includegraphics[width=\textwidth]{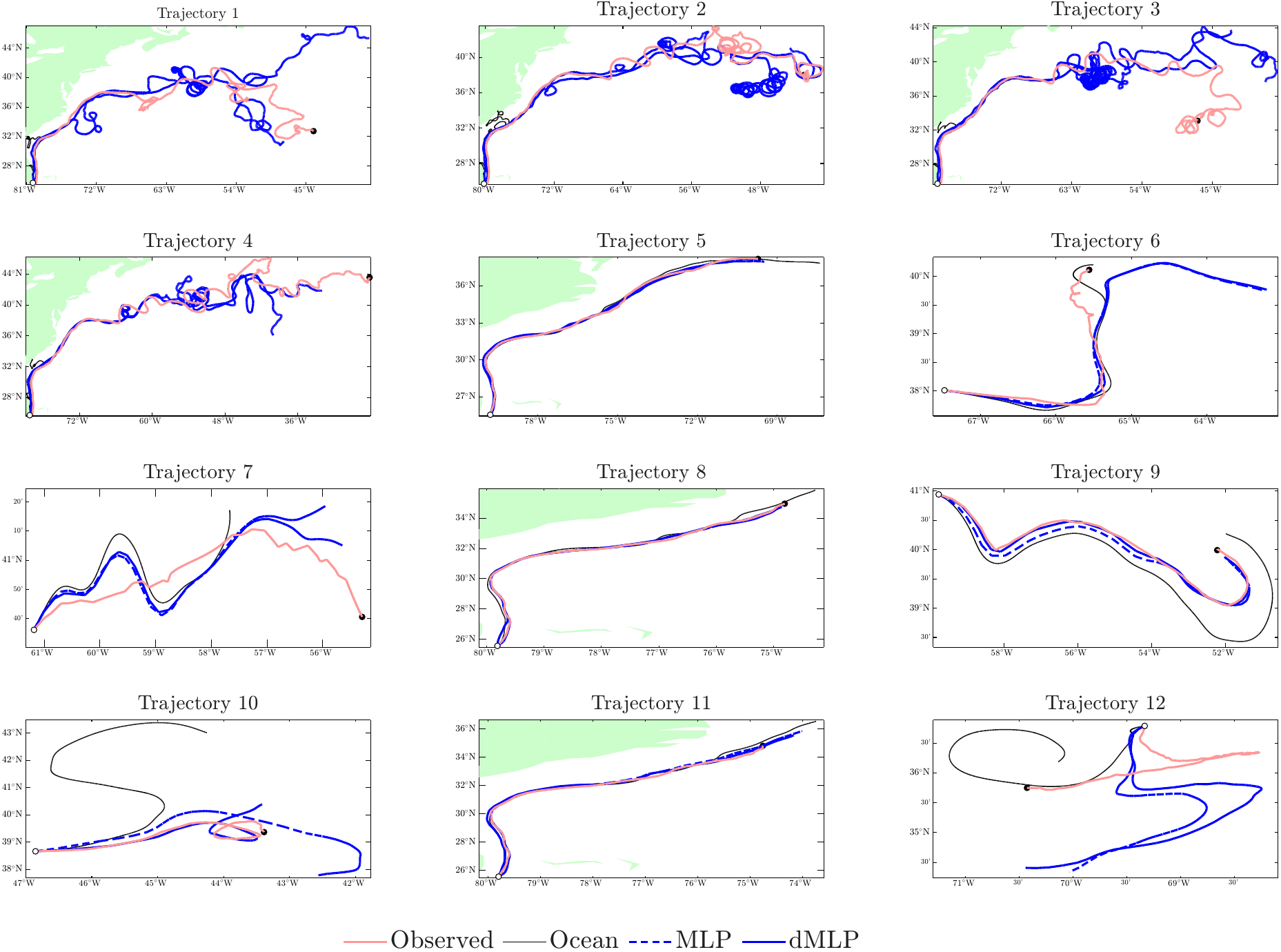}
    \caption{
    Individual leave-one-trajectory-out MLP trajectory comparisons for the Gulf Stream application. Observed ``Sargassum-following'' drifter trajectories are compared with the AVISO ocean baseline, the instantaneous MLP correction, and the delayed MLP correction. The instantaneous MLP is shown as a dashed blue curve and the delayed MLP as a solid blue curve. Filled circles denote trajectory initiation and open circles denote observed terminal locations.
    }
    \label{fig:loto_individual_gulfstream}
\end{figure}

Figure~\ref{fig:loto_individual_gulfstream} shows individual LOTO trajectory comparisons for both predictive models. Observations correspond to ``Sargassum-following'' drifter trajectories, ocean trajectories are obtained from the AVISO baseline circulation product, dashed blue curves denote the instantaneous MLP, and solid blue curves denote the delayed MLP. Aggregate statistics partially obscure substantial trajectory-to-trajectory variability. Several withheld trajectories are substantially improved by the instantaneous MLP alone, whereas delayed MLP can provide stronger corrections in some coherent jet-following pathways. Other cases remain difficult for both MLP and dMLP, especially in recirculatory or extrapolative regimes. The delayed MLP therefore acts primarily as a finite-memory extension of an already effective predictive closure rather than as a uniformly superior model.

\begin{figure}[t!]
    \centering
    \includegraphics[width=.75\textwidth]{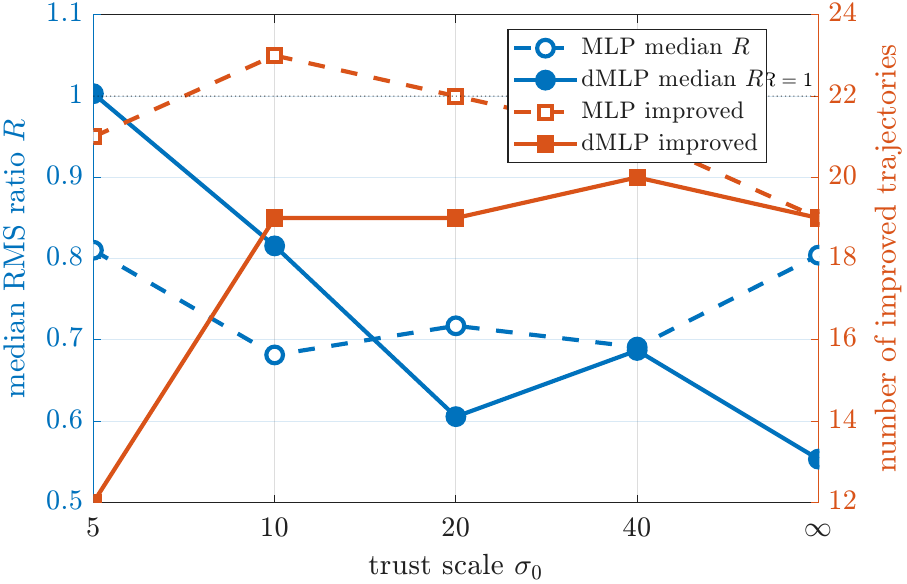}
    \caption{
    Trust-scale sensitivity for the instantaneous MLP and delayed MLP ensembles in the Gulf Stream application. Curves show the median trajectory RMS ratio and the number of withheld trajectories improved relative to the AVISO ocean baseline as functions of the trust scale $\sigma_0$. The instantaneous MLP is shown with open/dashed markers and the delayed MLP with filled/solid markers.
    }
    \label{fig:sigma0_sweep_gulfstream}
\end{figure}

The predictive corrections are modulated through an ensemble-based trust factor intended to suppress extrapolative corrections in poorly constrained regions of state space. Figure~\ref{fig:sigma0_sweep_gulfstream} summarizes sensitivity to the trust scale $\sigma_0$ for both MLP and dMLP. Small values suppress learned corrections strongly and keep trajectories close to the ocean baseline. The instantaneous MLP remains comparatively robust under strong trust suppression, whereas delayed MLP becomes more effective at moderate and weak suppression levels where longer coherent jet-following structure can be exploited. This increased flexibility, however, comes with greater trajectory-to-trajectory variability.

\begin{figure}[t!]
    \centering
    \includegraphics[width=.75\textwidth]{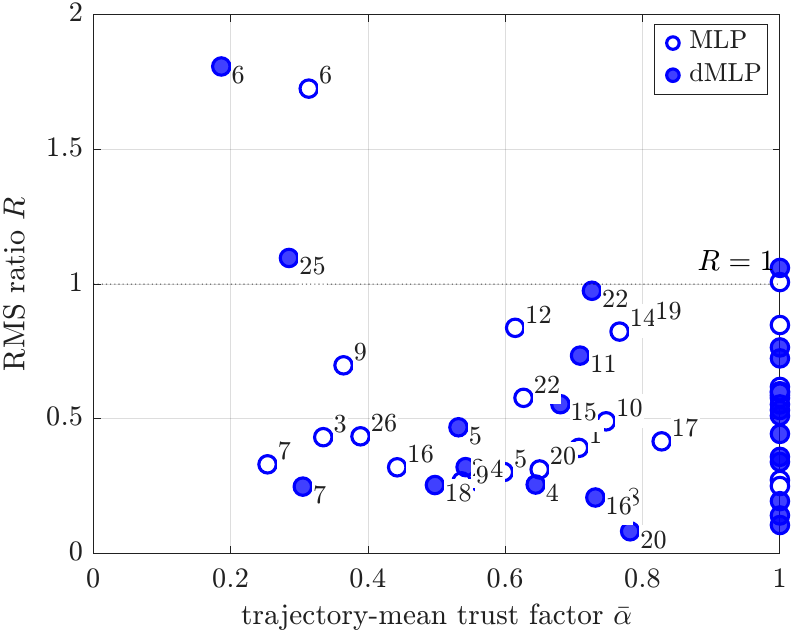}
    \caption{
    Trust--performance diagnostic for the instantaneous MLP and delayed MLP ensembles in the Gulf Stream application. Each point corresponds to one withheld trajectory segment. The horizontal axis is the trajectory-mean trust factor $\bar\alpha$ and the vertical axis is the trajectory RMS ratio relative to the AVISO ocean baseline. Open circles denote the instantaneous MLP and filled circles denote the delayed MLP.
    }
    \label{fig:uncertainty_trust_scatter_gulfstream}
\end{figure}

Figure~\ref{fig:uncertainty_trust_scatter_gulfstream} compares trajectory performance with the trajectory-mean trust factor. Open circles denote the instantaneous MLP and filled circles denote the delayed MLP. Ensemble spread provides a useful uncertainty proxy, but strong ensemble agreement alone does not guarantee accurate prediction. Several improved trajectories occur at moderate trust levels, whereas more strongly suppressed trajectories are often associated with degraded predictive regimes. In the Gulf Stream application, trust modulation is particularly important for delayed predictive corrections because delayed MLP exhibits greater sensitivity across coherent jet-following and extrapolative regimes. The trust factor acts primarily as an extrapolation-control diagnostic rather than a direct error estimator.

The sparse symbolic results differ substantially from the predictive neural models. Table~\ref{tab:modelsummary_gulfstream} shows that instantaneous SINDy provides the dynamically meaningful sparse correction for the Gulf Stream application, whereas delayed SINDy does not improve out-of-sample trajectory prediction when coupled back into trajectory integration. The Gulf Stream symbolic results therefore emphasize instantaneous SINDy as the effective local symbolic closure.

Although delayed states admit a direct physical interpretation in the Gulf Stream application, with one-step and two-step delays corresponding to 6-hour and 12-hour lags, respectively, the delayed sparse symbolic branch does not yield an improved trajectory closure. Bootstrap analysis nevertheless reveals persistent delayed algebraic structure involving ocean-velocity memory, atmospheric forcing, and short-time delayed coupling.

The instantaneous sparse symbolic correction takes the form
\begin{align}
    \dot x
    &=
    u
    +
    \widetilde G_x(\varphi),
    \\
    \dot y
    &=
    v
    +
    \widetilde G_y(\varphi).
\end{align}
A representative instantaneous sparse structure inferred from the LOTO analysis is
\begin{align}
    \widetilde G_x
    \approx\;&
    -0.83\,u
    +
    0.31\,u_a
    -
    1.42\,\omega
    -
    2.11\,\nabla\!\cdot\mathbf u,
    \\
    \widetilde G_y
    \approx\;&
    -0.76\,v
    +
    0.28\,v_a
    +
    1.35\,\omega
    -
    2.47\,\nabla\!\cdot\mathbf u.
\end{align}
These representative coefficients highlight the compact local structure recovered by instantaneous SINDy. Although coefficients vary across withheld trajectories, the retained terms repeatedly involve ocean velocity, atmospheric forcing, and rotational or divergence diagnostics, consistent with interpretable local transport corrections.

Delayed SINDy produces persistent lag-heavy algebraic structure, but this structure does not translate into a useful delayed trajectory closure. Bootstrap resampling shows that many delayed sparse symbolic terms are repeatedly selected across resampled datasets. The delayed sparse structure is dominated by ocean-velocity memory, atmospheric forcing, acceleration diagnostics, and short-time delayed velocity coupling across both 6-hour and 12-hour lags. The recurrent delayed dependencies indicate persistent short-memory structure across bootstrap realizations, even though no useful delayed sparse symbolic closure emerges for Gulf Stream trajectory prediction.

\begin{table}[t!]
    \centering
    \begin{tabular}{lclc}
    \toprule
    $x$-equation term & Frequency & $y$-equation term & Frequency \\
    \midrule
    $u$ & 0.884 & $v$ & 0.982 \\
    $u_{n-1}$ & 0.872 & $v_{n-1}$ & 0.978 \\
    $u_{n-2}$ & 0.861 & $v_{n-2}$ & 0.971 \\
    $\omega$ & 0.846 & $\omega$ & 0.963 \\
    $u_a$ & 0.831 & $v_a$ & 0.958 \\
    $(u_a)_{n-1}$ & 0.817 & $(v_a)_{n-1}$ & 0.947 \\
    $(u_a)_{n-2}$ & 0.804 & $(v_a)_{n-2}$ & 0.934 \\
    $D_tu$ & 0.791 & $D_tv$ & 0.921 \\
    $(D_tu)_{n-1}$ & 0.776 & $(D_tv)_{n-1}$ & 0.906 \\
    $\nabla\!\cdot\mathbf u$ & 0.765 & $\nabla\!\cdot\mathbf u$ & 0.894 \\
    \bottomrule
    \end{tabular}
    \caption{
    Most persistent delayed SINDy terms from bootstrap resampling for the Gulf Stream application. Subscripts denote fixed-time delays relative to sampling index $n$ with $\Delta t=6$ hr, so that $n-1$ and $n-2$ correspond to 6-hour and 12-hour lags, respectively. Frequencies denote the fraction of bootstrap fits in which a term remains active.
    }
    \label{tab:bootstrap_gulfstream}
\end{table}

\begin{figure}[t!]
    \centering
    \includegraphics[width=\textwidth]{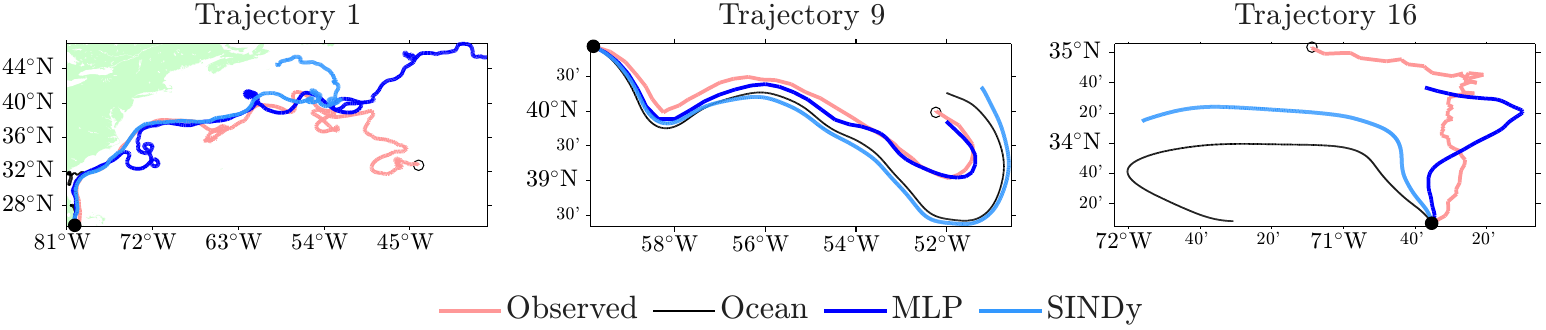}
    \caption{
    Representative leave-one-trajectory-out trajectory comparisons for the Gulf Stream application. Observed ``Sargassum-following'' drifter trajectories are compared with the AVISO ocean baseline, the instantaneous MLP correction, and the instantaneous sparse symbolic SINDy correction. Filled circles and open circles indicate trajectory start and end points.
    }
    \label{fig:trajectory_examples_gulfstream}
\end{figure}

Figure~\ref{fig:trajectory_examples_gulfstream} illustrates representative withheld trajectories used to compare the predictive and sparse symbolic models. For the Gulf Stream application, the predictive model shown is the instantaneous MLP, whereas the sparse symbolic model is instantaneous SINDy. The examples highlight the different behavior of the predictive and sparse symbolic closures. MLP provides stronger trajectory-level corrections, while SINDy provides smoother and more conservative symbolic transport corrections. Trajectory~1 illustrates a case in which the ocean baseline already captures the initial jet-following motion reasonably well, while MLP later develops a spurious loop and SINDy remains closer to the baseline transport. Trajectory~9 provides the clearest MLP improvement together with a conservative symbolic correction. Trajectory~16 illustrates a difficult extrapolative regime in which all closures struggle.

The Gulf Stream experiments show that instantaneous diagnostics already contain substantial predictive structure. A simple residual MLP therefore already produces strong trajectory corrections. Delayed states remain useful, but their role is more selective and depends strongly on the closure itself. Delayed MLP can improve some coherent jet-following regimes, although with greater sensitivity to trust modulation and extrapolative instability. The sparse symbolic results lead to a different conclusion: instantaneous SINDy remains dynamically meaningful, whereas delayed sparse symbolic structure, despite persistent delayed terms under bootstrap resampling, does not produce a useful delayed trajectory closure. Overall, the Gulf Stream results indicate that predictive transport structure is already strongly encoded within local ocean--atmosphere diagnostics, while delayed structure plays a more selective role associated primarily with coherent jet-following regimes and delayed algebraic transport structure.

\section{Conclusions}

This work develops a data-driven approach for learning effective transport corrections from sparse Sargassum-following drifter trajectories using physically guided instantaneous and delayed ocean--atmosphere diagnostics. The methodology combines baseline circulation models, leave-one-trajectory-out validation, predictive residual learning, sparse symbolic discovery, and bootstrap stability analysis within a common delayed-transport setting motivated in part by finite-memory effects in inertial-particle transport.

The Puerto Rico and Gulf Stream applications support a consistent overall picture. In Puerto Rico, delayed sparse symbolic corrections remain modestly but systematically effective under a fixed-time delayed formulation, supporting weak but recurrent short-memory transport structure. In the Gulf Stream application, instantaneous sparse symbolic corrections remain dynamically meaningful, whereas delayed sparse symbolic closure does not improve out-of-sample trajectory evolution despite persistent delayed-term selection under bootstrap resampling.

Instantaneous MLP already captures substantial predictive structure in both applications, while delayed MLP can provide additional improvement in some coherent transport regimes but with greater sensitivity and stronger trajectory dependence. The results therefore indicate that physically guided local ocean--atmosphere diagnostics already contain substantial predictive transport structure before delayed memory is introduced.

The sparse symbolic results lead to a different conclusion. In Puerto Rico, delayed SINDy identifies weak but recurrent delayed symbolic structure associated with short-time velocity memory and ocean--atmosphere forcing. In the Gulf Stream application, delayed symbolic structure persists algebraically under bootstrap resampling but does not yield a useful delayed trajectory closure when coupled back into transport evolution. Instantaneous SINDy nevertheless remains dynamically meaningful through compact local symbolic corrections involving velocity, atmospheric forcing, and rotational or divergence diagnostics.

Overall, the results suggest that delayed transport states can reveal physically interpretable short-memory structure even when delayed predictive improvement remains modest or strongly trajectory dependent. The learned corrections should nevertheless be viewed as effective coarse-grained transport closures rather than complete mechanistic transport laws, since the available observations remain sparse, unresolved transport processes are important, and ocean--atmosphere forcing products themselves contain uncertainty.

Future work will investigate whether larger ensembles of controlled macroscopic transport trajectories can clarify when coarse-grained transport dynamics generate effective finite-memory structure and when such delayed predictive information admits stable sparse symbolic closure.

\section*{Acknowledgments}

Cedric Gigand designed the Sargassum-following drifters used in the Gulf Stream deployments, and Cedric Gigand, Gustavo Goni, and Gage Bonner assisted with the drifter deployments. The authors thank Gage Bonner for reading the manuscript and useful suggestions.

\section*{Funding}

This research was funded by the National Science Foundation (NSF) Grant No. OCE2148499.

\section*{Author Declarations}

\subsection*{Conflict of Interest}

The authors have no conflict of interest to disclose.

\section*{Author contributions}

F.J.B.-V.\ conceived the study, developed the theoretical framework, performed the numerical analysis, developed the MATLAB implementation, generated the figures, and wrote the manuscript. M.J.O.\ designed the Gulf Stream Sargassum drifter campaign, acquired funding together with F.J.B.-V., and contributed to manuscript editing and revision. J.M.\ coordinated the FVCOM implementation, acquired funding, and contributed to manuscript editing and revision. E.C.\ designed the Puerto Rico Sargassum drifter campaign and contributed to manuscript editing and revision. ChatGPT assisted with code development, debugging, and manuscript editing.

\bibliographystyle{alpha}
\bibliography{fot}

\section*{Data and Software Availability}

The FVCOM implementation used in this study was produced by the Caribbean Coastal Ocean Observing System (CARICOOS) and is available from \href{https://dm1.caricoos.org/thredds/catalog/content/FVCOM/PRVI/catalog.html}{https://dm1.caricoos.org/thredds/catalog/content/FVCOM/PRVI/catalog.html}.

The gridded multimission altimeter products were produced by SSALTO/DUACS and distributed by AVISO (\href{https://www.aviso.altimetry.fr/}{https://www.aviso.altimetry.fr/}) with support from CNES.

The wind velocity data were obtained from the European Centre for Medium-Range Weather Forecasts (ECMWF) Reanalysis v5 (ERA5), available at \href{https://www.ecmwf.int/en/forecasts/dataset/ecmwf-reanalysis-v5}{https://www.ecmwf.int/en/forecasts/dataset/ecmwf-reanalysis-v5}.

Puerto Rico and Gulf Stream drifter trajectory data used in this study are available from the corresponding author upon reasonable request.

The MATLAB scripts used to generate the figures and results presented in this paper are available from the corresponding author upon reasonable request.

\end{document}